# Superconductivity in La$_{1-x}$Sm$_x$O$_{0.5}$F$_{0.5}$BiS$_2$ ($x$ = 0.2, 0.8) under hydrostatic pressure

G. Kalai Selvan[1‡], Gohil Thakur[2‡], K. Manikandan[1], A. Banerjee[3], Zeba Haque[2], L. C. Gupta[2†], Ashok Ganguli[2] and S. Arumugam[1*]

[1]*Centre for High Pressure Research, School of Physics, Bharathidasan University, Tiruchirapalli 620024, India.*

[2]*Solid State and Nano Research laboratory, Department of Chemistry, Indian Institute of Technology New Delhi 110016, India.*

[3]*UGC-DAE Consortium for Scientific Research, University Campus, Khandwa Road, Indore-452001, India.*

**Abstract**

We have investigated the pressure effect on the newly discovered samarium doped La$_{1-x}$Sm$_x$O$_{0.5}$F$_{0.5}$BiS$_2$ superconductors. More than threefold increase in $T_c$ (10.3 K) is observed with external pressure (at ~1.74 GPa at a rate of 4.08 K/GPa)) for $x$ = 0.2 composition. There is a concomitant large improvement in the quality of the superconducting transition. Beyond this pressure $T_c$ decreases monotonously at the rate of -2.09 K/GPa. In the $x$ = 0.8 sample, we do not observe any enhancement in $T_c$ with application of pressure (up to 1.76 GPa). The semiconducting behavior observed in the normal state resistivity of both of the samples is significantly subdued with the application of pressure which, if interpreted invoking thermal activation process, implies that the activation energy gap of the carriers is significantly reduced with pressure. We believe these observations should generate further interest in the La$_{1-x}$Sm$_x$O$_{0.5}$F$_{0.5}$BiS$_2$ superconductors.

***Key Words*:** La$_{1-x}$Sm$_x$O$_{0.5}$F$_{0.5}$B$_2$; pressure effect on $T_c$; activation energy gap

[‡] These authors contributed equally
**Corresponding Author:** Email: - [*]sarumugam1963@yahoo.com
Telephone no: - (O): **91-431- 2407118 Cell: 91-95009 10310;**
Fax No: - **91-431- 2407045, 2407032**

## 1. Introduction

In year 2008, discovery, of new superconducting compounds like Fe based pnictides LnFeAsO$_{1-x}$F$_x$ with $T_c$ exceeding 50 K has renewed interest in superconductivity in correlated electron systems in condensed matter physics[1–7]. Recently superconductivity has been reported in the layered compound Bi$_4$O$_4$S$_3$, $T_c$ = 8.6 K[8,9]. Following this report, other BiS$_2$-based superconductors including electron doped LnO$_{1-x}$F$_x$BiS$_2$ (Ln = La, Ce, Pr, Nd, Yb)[10–18], La$_{1-x}$M$_x$OBiS$_2$ (M = Ti, Zr, Hf, Th)[19] and Sr$_{1-x}$Ln$_x$FBiS$_2$[20–22] have been found with $T_c$ ~ 2.5 K to 10 K. It may be pointed out that Sm-based compound SmO$_{1-x}$F$_x$BiS$_2$ doesn't exhibit superconductivity down to 2 K[23], while the other rare earth compounds (Ln = La, Ce, Pr, Nd and Yb) are superconducting (with, $T_c$ = 2.5 – 5.6 K)[10,13–15,17,18,24,25]. These compounds share a common BiS$_2$ layer which serves as the basic structural building block of this new superconducting family. As shown in figure 1 the structure consists of a stacking of alternating BiS$_2$ bilayers and Ln-O spacer layers. Electrical resistivity under pressure[21,26–30] revealed that $T_c$ of BiS$_2$ based superconductors is sensitive to application of pressure and can be significantly enhanced as observed in the cuprates[31,32] and Fe-based families[6,7] The structural changes caused by the application of external hydrostatic pressure results in a suppression of the semiconducting behavior and enhancement in $T_c$ in LaO$_{0.5}$F$_{0.5}$BiS$_2$ to 10.1 K (6.1 K/GPa)[26,27,30,33], 6.7 K for CeO$_{0.5}$F$_{0.5}$BiS$_2$ (2.2 K/GPa)[34], PrO$_{0.5}$F$_{0.5}$BiS$_2$ to 7.6 K (2.5 K/GPa)[34] and NdO$_{0.5}$F$_{0.5}$BiS$_2$ to 6.4 K (1.56 K/GPa)[34] and on further increasing the pressure $T_c$ decreases slowly in the same way as in the layered pnictides and chalcogenides. A general trend is evident, that at ambient pressure substitution of smaller rare earth ion increases the $T_c$ and on the other hand the effect of pressure seems to diminish with the size of rare earth in LnOBiS$_2$[25,33]. In the normal state of these materials, there is a significant suppression of semiconducting behavior with pressure which is continuous up to the transition pressure 'Pt' (Pt = Pressure induced semiconductor-metal transition). A rapid increase of the charge carrier density is inferred from both the suppression of the semiconducting behavior and the rapid increase of $T_c$ in this region. With increasing pressure the BiS$_2$-type compounds show an abrupt transformation from the low $T_c$ phase to a high $T_c$ one, with a characteristic maximum pressure ranging from ~ 1.5 to 2.5 GPa [18,26,34]. In order to achieve an optimum chemical doping it is essential to understand the pressure effect on superconductivity and suppression of the energy gap in this material. Here, we report the variation of $T_c$ of the La$_{1-x}$Sm$_x$O$_{0.5}$F$_{0.5}$BiS$_2$ ($x$ = 0.2, 0.8) samples at ambient and under external hydrostatic pressure studied with dc magnetic susceptibility (0.8 GPa) and electrical resistivity measurements (~ 2 GPa). Our results show enhancement from $T_c$ = 3.2 K at ambient pressure to $T_c$ upto ~ 10.3 K at an applied pressure 1.74 GPa in $x$ = 0.2. However, $T_c$ for $x$ = 0.8 sample exhibits almost no change with application of pressure. Semiconducting and metallic behavior of La$_{1-x}$Sm$_x$O$_{0.5}$F$_{0.5}$BiS$_2$ has been explained with the help of energy gap.

## 2. Experimental details

Polycrystalline samples La$_{1-x}$Sm$_x$O$_{0.5}$F$_{0.5}$BiS$_2$ ($x$ = 0, 0.2 and 0.8) studied were synthesized by solid state reaction route and characterized as described by Thakur *et al*[23]. The dc magnetization studies at various applied pressures up to ~1 GPa were performed down to 2 K in an applied magnetic field of 10 Oe in zero-field cooling with a vibrating sample magnetometer (VSM) option in the Physical Property Measurement System (PPMS, Quantum Design, USA). The external pressure was generated by a piston-cylinder pressure cell (Easy Lab Technologies Mcell 10). The applied pressure in the cell was calculated by the change of superconducting transition temperature of pure Sn (99.99%). A mixture of Fluorinert FC 70 and FC 77 (1:1) was used as pressure transmitting medium. The normalized (dc) electrical resistivity measurements under ambient and various hydrostatic pressures (up to ~ 2 GPa) down to 4K were carried out by conventional four-probe method using clamp type hybrid hydrostatic pressure cell and Closed Cycle Refrigerator - Variable Temperature Insert (CCR-VTI) system. The sample size of ~ 1.0×0.8×0.4 mm$^3$ was used in these experiments. For conventional four probe method, the contacts were made by high quality silver paste with copper wire of Φ 0.15 mm and the Daphne #7474 was used as a pressure transmitting medium and the actual pressure inside the cell was calculated using a calibration curve that was previously obtained from structural transitions of bismuth at room temperature by *Honda et al*[35].

## 3. Results and Discussion

We reported enhanced superconductivity in La$_{1-x}$Sm$_x$O$_{0.5}$F$_{0.5}$BiS$_2$ system by means of chemical pressure, replacing larger La ions by smaller Sm ions. We found a $T_c$ increases upon substituting Sm ions for La; $T_c$ = 3.2 K for $x$ = 0.2 and a maximum $T_c$ of 4.6 K for $x$ = 0.8 composition[23]. Here, we have chosen lower ($x$ = 0.2) and higher ($x$ = 0.8) samarium doped samples of $La_{1-x}Sm_xO_{0.5}F_{0.5}BiS_2$ series for our high pressure studies.

### 3.1 Magnetic measurements

Magnetization measurement in zero field cooled condition at low field (H =10 Oe) in the temperature range of 2 – 10 K for the samples $x$ = 0.2 and 0.8 are shown in figure 2 (a and b). As observed earlier[23] a clear onset of diamagnetic signal corresponding to superconductivity with $T_c$ ~ 2.8 K was observed for $x$ = 0.2 (Fig.2 (a)) and at 4.3 K for $x$ = 0.8 (Fig.2 (b)) at ambient pressure. On application of pressure this diamagnetic signal shift smoothly to higher temperature for $x$ = 0.2 and also the superconducting volume fraction increases. A maximum $T_c$ of 6 K is seen in sample with $x$ = 0.2 at a pressure of 0.8 GPa (limit of our magnetization pressure cell). The variation of $T_c$ with respect to pressure i.e. the rate of pressure coefficient for $x$ = 0.2 sample is ~3.78 K/GPa, which is quite large as compared to other rare-earth based systems Ln(O,F)BiS$_2$[33,34]. Whereas no change either in $T_c$ (~4.3 K) or in the superconducting volume fraction is observed in the sample with $x$ = 0.8. Importantly, this is the first report of pressure dependent magnetization studies in LnOBiS$_2$ materials.

### 3.2 Resistivity measurements

Temperature dependence of resistivity for $La_{1-x}Sm_xO_{0.5}F_{0.5}BiS_2$ ($x$ = 0.2 and 0.8) at various applied pressures are shown in the Fig. 3. Superconductivity at ambient pressure is characterized by a sharp drop in resistivity to zero at 3.2 K for $x$ = 0.2 and 4.6 K for $x$ = 0.8 samples (insets of fig.3. (a and c)). The onset $T_c$ is determined as an intersection of two extrapolated lines drawn through 90% and 10% resistivity transition curve (as indicated in the insets of fig.3.(a and c)). This significant enhancement in $T_c$ upon increasing Sm doping has been attributed to the increase in chemical pressure generated due to substitution of smaller Sm ions at the La site[23,36,37]. On application of pressure there is a significant increase in $T_c$ onset from 3.2 K at P = 0 GPa to 10.3 K at P = 1.76 for $x$ = 0.2 sample, as seen in figure 3b. It is to be noted that a zero resistance state for $x$ = 0.2 sample is not observed in figure 3 (a and b) since our CCR-VTI system were limited to 3.5 K and therefore the ambient pressure resistivity data of $x$ = 0.2 showing the superconducting transition is plotted in the inset of figure 3a. At pressure ~0.35 GPa and 0.62 GPa the superconducting transition is broad with $T_c$ onset ~ 6.9 K and 7.4 K respectively ($T_c$ zero not seen till 3.5 K). A sharp transition is however observed at higher pressure (upto 1.76 GPa) with maximum $T_c$ onset 10.3 K and $T_c^{zero}$ to 9.2 K (at 1.74 GPa). The highest pressure coefficient $dT_c/dP$ for $x = 0.2$ (La$_{0.8}$Sm$_{0.2}$O$_{0.5}$F$_{0.5}$BiS$_2$) is estimated to be around ~4.1 K/GPa at 1.74 GPa. On further increasing pressure $T_c^{onset}$ suddenly decreases to 6.2 K (at 1.96 GPa) with a pressure coefficient of -2.09 K/GPa. The variation of $T_c^{onset}$ from resistivity data as a function of pressure for $x = 0.2$ is plotted in figure 3(e). $T_c$ values obtained in the magnetization measurements are slightly lower than what obtained from resistivity studies due to the inhomogeneity of the polycrystalline sample. The sharp increase in superconducting transition temperature under hydrostatic pressure of just above 1 GPa may be due to structural phase transformation as observed in other recently studied systems[27,38]. Also, there is a possibility of the strong electron correlation in BiS$_2$ superconducting layers being enhanced under pressure.

Figure 3 (c and d) shows the plots of temperature dependence of the electrical resistivity ρ(T) for sample with $x = 0.8$ composition at various applied pressures. From figure 3(d and f) it is clearly seen that with pressure the $T_c$ onset increases insignificantly from ~4.6 K to 4.93 K (at 1.76 GPa). $T_c(\rho=0)$ also increases very slightly from around 4.1 K to 4.4 K. The estimated $dT_c^{onset}/dP$ for $x = 0.8$ is ~0.028 K/GPa. Resistivity data corroborates the magnetization data where no significant enhancement in $T_c$ was observed (figure 3 (f)).

Semiconducting behavior in the normal state resistivity is seen at various pressures for $x = 0.2$ and $x = 0.8$. While this semiconducting behavior is suppressed drastically with increase in pressure for $x = 0.2$ sample, only slight improvement in metallicity was observed for $x = 0.8$ sample (figure 3 (a and c)). At pressure $P = 1.96$ GPa the sample with $x = 0.2$ becomes metallic. The normal state resistivity also keeps on decreasing with applied pressures up to 1.74 GPa for both the sample with $x = 0.2$ and 0.8.

To explore the normal state semiconducting properties and to see their trend with pressure, we show, in figure 4, the log(ρ) vs 1/T plot for $x = 0.2$ and 0.8 samples at various pressures up to 1.96 GPa for $x = 0.2$ and 0.8. We do this considering that the resistivity in the semi-metallic regime is governed by thermal activation of carriers and the values of energy gaps may be determined from these plots. The observed resistivity versus temperature data in the entire temperature range could not be fitted for a single gap. The observed resistivity data could be described in two different temperature regions by the following activation energy relation.

$$\rho(T)=\rho_0 e^{\Delta(1,2)/2K_BT},$$

where, $\rho_0$ is constant, $\Delta$ is energy gap and $k_B$ is the Boltzmann constant. The fitting has been done in two distinct regions, in the temperature range 300 to 200 K with energy gap $\Delta_1$ (vertical linear fit), and in the temperature range ~25 K to $T_c^{onset}$ with energy gap $\Delta_2$ (horizontal linear fit). Analysis of $\rho(T)$ data at ambient pressure gives the estimated values of the energy gaps $\Delta_1/k_B$ ~ 27.45 K and $\Delta_2/k_B$ ~ 1.19 K for the $x = 0.2$ compound. Interestingly, these values obtained are considerably less than those reported for other rare earth based $BiS_2$ based superconductors [26,27,30,34,38,39]. This is due to improved normal state properties of $La_{0.8}Sm_{0.2}O_{0.5}F_{0.5}BiS_2$. Further we show in figure 5 the variation of $\rho$ at 11 K just above the $T_c$, Residual Resistivity Ration (*RRR*) and energy gaps ($\Delta_1/k_B$, $\Delta_2/k_B$) with various applied pressure from 0 to 1.96 GPa for $x = 0.2$ and 0.8. In sample with $x = 0.2$, (figure 5 a and b) the rise in resistivity at low temperatures induced by the semiconducting behavior is markedly suppressed upon applying pressure, particularly beyond 1 GPa. It can be clearly seen from Fig. 5(b) that both energy gaps $\Delta_1$ and $\Delta_2$ (for $x = 0.2$) decreases gradually with applied pressure till 1.5 GPa. The decrease is quite rapid beyond this pressure. The decrease of energy gap with increasing pressure, suggests an increase of charge carriers density at the Fermi surface with application of pressure. Interestingly, first principal calculations have suggested that there could be either insulator/semiconductor to metal transition accompanied with superconductivity at low temperature under pressure[40,41]. As seen in figure 3 (e) the onset $T_c$ initially shows a steep increase for $x = 0.2$ sample; however, it gradually decreases above ~ 1.7 GPa. The rate change of $T_c$ above 1.7 GPa is ~ -2.09 K/GPa. The observation of highest $T_c$ achieved coincides with sudden decrease in the $\Delta/k_B$ values indicating a possible phase transformation. The highest $T_c$ obtained at 1.74 GPa is achieved with a marked suppression of the semiconducting behavior, and $T_c$ then suddenly decreases upon approaching the metallic region when both the energy gaps overlap. Clear observations of decreasing semiconducting behavior in the normal state resistivity is observed as also indicated by large decrease in $\rho^{11K}$ values, ($\rho^{11K} = 0.15$ *mΩ-cm* at 0 GPa, and 0.07 $m\Omega$-cm at 1.74 GPa and 0.009 $m\Omega$-cm at 1.96 GPa). This is evident in figure 5 a. The decrease in normal state resistivity ($\rho^{11K}$), by more than an order of magnitude at a pressure of 0.35 GPa along with the change from semiconducting to metallic behavior is quite interesting. It is believed that under pressure S-Bi-S bond angle along with inter-atomic distances change, which in turn affects the charge density at Fermi surface and hence a clear change in normal state electrical transport properties.

Figure 5 (c and d) shows the estimated $\rho$ at 11 K just above the $T_c$ and Residual Resistivity Ration (RRR), and variation of calculated energy gaps ($\Delta_1/k_B$, $\Delta_2/k_B$) with applied pressure from 0 to 1.76 GPa for sample with $x = 0.8$. Normal state properties of $x = 0.8$ composition do not change much. Thus the variation in the energy gaps $\Delta_1/k_B$ and $\Delta_2/k_B$ with the applied pressure to 1.76 GPa is also not very prominent. The magnitude of $\rho^{11K}$ decrease from 0.03 $m\Omega$-cm to 0.02 $m\Omega$-cm at 1.76 GPa. There is an insignificant variation in the $T_c$ ($T_c$ 4.58 K at 0 GPa and 4.62 K at 1.76 GPa) with a rate of pressure coefficient 0.028 K/GPa (figure 2 f). It is observed that with increasing pressure the normal state semiconducting behavior also does not change much. This is slightly different quantitatively from earlier reports on pressure dependent superconductivity of other BiS$_2$ based superconducting (La/Pr/Nd/Ce)O$_{0.5}$F$_{0.5}$BiS$_2$ compounds[26–28,30,34,42].

The coexistence of insulator to metal transition and threefold increase of $T_c$ in $x = 0.2$ on one hand and no change in $T_c$ and normal state properties in $x = 0.8$ under applied pressure on the other hand warrants further detailed structural studies under pressure in these new materials. A qualitative explanation of different behavior of these two Sm-doped samples is discussed below. In La$_{1-x}$Sm$_x$O$_{0.5}$F$_{0.5}$BiS$_2$ substitution of small amount of Sm at La sites produces a small internal chemical pressure by lattice contraction and slightly enhances the effective overlap of Bi and S orbital[23]. This lattice contraction is further facilitated by external pressure. As the Fermi-level is sensitive to structural compression, pressure would lead to upshift of Fermi level[43] and hence a high $T_c$ (~ 10 K) is achieved. In case of $x = 0.8$ which already has a relatively high $T_c$ (4.6 K) at ambient pressure, the lattice contraction is already too strong and hence applying external pressure would only lead to excess of the charge carrier density in the system in which case pressure would have no effect on $T_c$.

4## 4. Conclusion

We report here how superconductivity in the new superconductors La$_{1-x}$Sm$_x$O$_{0.5}$F$_{0.5}$BiS$_2$ ($x$ = 0.2, 0.8) is influenced by the application of hydrostatic pressure. We show that superconducting transition temperature $T_c$ in La$_{0.8}$Sm$_{0.2}$O$_{0.5}$F$_{0.5}$BiS$_2$ is increased from 3.2 K to above 10.3 K under pressure just above ~1.5 GPa. This is a dramatic more than threefold enhancement of $T_c$. The quality of superconducting transition is also very significantly improved under high pressures. In addition there is a concomitant improvement in the normal state resistance and suppression of semi-metallic behavior of the material. While there is virtually no effect of pressure on $T_c$ of the $x$ = 0.8 material, there does occur a transformation from semiconductor to metallic behavior in the normal state just as in the sample with $x$ = 0.2. The precise mechanism driving the enhancement of $T_c$ with pressure is not clear at present. High pressure X-ray diffraction experiments on the samples investigated here are needed to determine whether the pressure-induced enhancement of $T_c$ and the suppression of semiconducting behavior in the normal state are related to a structural transition.

**Acknowledgement**

The author SA thank DST, (SERB, TSDP, FIST), BRNS, DRDO and CEFIPRA for the financial support. AKG thanks SERB (DST) for funding. GKS thanksUGC- BSR- RFSMS- SRF for the meritorious fellowship. GST thanks CSIR (Govt. of India) for a fellowship. The authors thank Mr. Kranti Kumar, UGC-DAE –CSR (Indore), who has helped to carry out magnetization measurements. SA and GK Selvan acknowledge UGC-DAE –CSR (Indore) for providing the user facilities for the measurements.

**Notes and References**

[†] Visiting scientist at Solid State and Nano Research Laboratory, Department of Chemistry, IIT Delhi

[1]  Kamihara Y, Watanabe T, Hirano M and Hosono H 2008 Iron-Based Layered Superconductor LaO1-xFxFeAs (x = 0.05-0.12) with Tc = 26 K *J. Am. Chem. Soc.* **130** 3296–7

[2]  Chen G F, Li Z, Wu D, Li G, Hu W Z, Dong J, Zheng P, Luo J L and Wang N L 2008 Superconductivity at 41 K and its competition with spin-density-wave instability in layered CeO1-xFxFeAs *Phys. Rev. Lett.* **100** 1–4

[3]  Bos J-W G, Penny G B S, Rodgers J a, Sokolov D a, Huxley A D and Attfield J P 2008 High pressure synthesis of late rare earth RFeAs(O,F) superconductors; R = Tb and Dy. *Chem. Commun.* **003** 3634–5

[4]  Ren Z A, Lu W, Li Y, Yi W, Shen X-L, Li Z-C, Che G-C, Dong X-L, Sun L-L, Zhou F and Zhao Z-X 2008 Superconductivity at 55 K in Iron-Based F-Doped Layered Quaternary Compound Sm[O1-xFx]FeAs *Chinese Phys. Lett.* **25** 2215

[5]  Ren Z-A, Yang J, Lu W, Yi W, Shen X-L, Li Z-C, Che G-C, Dong X-L, Sun L-L, Zhou F and Zhao Z-X 2008 Superconductivity in the iron-based F-doped layered quaternary compound Nd[O1−x]FeAs *EPL* **82** 57002

[6]  Stewart G R 2011 Superconductivity in iron compounds *Rev. Mod. Phys.* **83** 1589–652

[7]  Ganguli A K, Prakash J and Thakur G S 2013 The iron-age of superconductivity: structural correlations and commonalities among the various families having -Fe-Pn- slabs (Pn = P, As and Sb). *Chem. Soc. Rev.* **42** 569–98

[8]  Mizuguchi Y, Fujihisa H, Gotoh Y, Suzuki K, Usui H, Kuroki K, Demura S, Takano Y, Izawa H and Miura O 2012 BiS2-based layered superconductor Bi4O4S3 *Phys. Rev. B* **86** 220510

[9]  Singh S K, Kumar A, Gahtori B, Kirtan S, Sharma G G, Patnaik S, Awana V S and Sharma G 2012 Bulk Superconductivity in Bismuth oxy-sulfide Bi4O4S3 Bulk Superconductivity in Bismuth oxy-sulfide Bi4O4S3 *J. Am. Chem. Soc.* **134** 16504–7

[10] Mizuguchi Y, Demura S, Deguchi K, Takano Y, Fujihisa H, Gotoh Y, Izawa H and Miura O 2012 Superconductivity in novel BiS2-based layered superconductor LaO1-xFxBiS2 *J. Phys. Soc. Jpn* **81** 114725

[11] Xing J, Li S, Ding X, Yang H and Wen H H 2012 Superconductivity appears in the vicinity of semiconducting-like behavior in CeO1-xFxBiS2 *Phys. Rev. B* **86** 214518

[12] Jha R and Awana V P S 2013 Superconductivity in Layered CeO0.5F0.5BiS2 *J. Supercond. Nov. Magn.* **27** 1–4

[13] Jha R, Kumar A, Singh S K and Awana V P S 2013 Superconductivity at 5K in NdO0.5F0.5BiS2 *J. Appl. Phys.* **113** 056102

[14] Yazici D, Huang K, White B D, Chang A H, Friedman A J and Maple M B 2013 Superconductivity of F-substituted LnOBiS2 (Ln=La, Ce, Pr, Nd, Yb) compounds *Philos. Mag.* **93** 673–80

[15] Demura S, Mizuguchi Y, Deguchi K, Okazaki H, Hara H, Watanabe T, Denholme S J, Fujioka M, Ozaki T, Fujihisa H, Gotoh Y, Miura O, Yamaguchi T, Takeya H and Takano Y 2013 BiS2 - based superconductivity in F-substituted NdOBiS2 *J. Phys. Soc. Jpn* **82** 033708

[16] Deguchi K, Mizuguchi Y, Demura S, Hara H, Watanabe T, Denholme S J, Fujioka M, Okazaki H, Ozaki T, Takeya H, Yamaguchi T, Miura O and Takano Y 2013 Evolution of superconductivity in LaO1−xFxBiS2 prepared by high-pressure technique *EPL* **101** 17004

[17] Jha R, Kumar A, Kumar Singh S and Awana V P S 2013 Synthesis and superconductivity of new BiS2 based superconductor PrO0.5F0.5BiS2 *J. Supercond. Nov. Magn.* **26** 499–502

[18] Yazici D, Jeon I, White B D and Maple M B 2015 Superconductivity in layered BiS2-based compounds *Phys. C* **514** 218–36

[19] Yazici D, Huang K, White B D, Jeon I, Burnett V W, Friedman a. J, Lum I K, Nallaiyan M, Spagna S and Maple M B 2013 Superconductivity induced by electron doping in La1−xMxOBiS2 (M= Ti, Zr, Hf, Th) *Phys. Rev. B* **87** 174512

[20] Lin X, Ni X, Chen B, Xu X, Yang X, Dai J, Li Y, Yang X, Luo Y, Tao Q, Cao G and Xu Z 2013 Superconductivity induced by la doping in Sr1-xLaxFBiS2 *Phys. Rev. B* **87** 020504(R)

[21] Jha R, Tiwari B and Awana V P S 2014 Impact of Hydrostatic Pressure on Superconductivity of Sr0.5La0.5FBiS2 *J phys. soc. Jpn* **83** 063707

[22] Li L, Li Y, Jin Y, Huang H, Chen B, Xu X, Dai J, Zhang L, Yang X, Zhai H, Cao G and Xu Z 2015 Coexistence of superconductivity and ferromagnetism in Sr 0 . 5 Ce 0 . 5 FBiS 2 *Phys. Rev. B* **014508** 014508

[23] Thakur G S, Selvan G K, Haque Z, Gupta L C, Samal S L, Arumugam S and Ganguli A K 2015 Synthesis and Properties of SmO0.5F0.5BiS2 and Enhancement in Tc in La1−ySmyO0.5F0.5BiS2 *Inorg. Chem.* **54** 1076–81

[24] Xing J, Li S, Ding X, Yang H and Wen H H 2012 Superconductivity appears in the vicinity of semiconducting-like behavior in CeO1−xFxBiS2 Jie *Phys. Rev. B* **86** 214518

[25] Mizuguchi Y 2014 Review of superconductivity in BiS2-based layered materials *J. Phys. Chem. Solids* **84** 34–48

[26] Wolowiec C T, Yazici D, White B D, Huang K and Maple M B 2013 Pressure-induced enhancement of superconductivity and suppression of semiconducting behavior in LnO0.5F0.5BiS2 (Ln=La,Ce) compounds *Phys. Rev. B* **88** 064503

[27] Tomita T, Ebata M, Soeda H, Takahashi H, Fujihisa H, Gotoh Y, Mizuguchi Y, Izawa H, Miura O, Demura S, Deguchi K and Takano Y 2014 Pressure-Induced Enhancement of Superconductivity and Structural Transition in BiS2 -Layered LaO1−xFxBiS2 *J. Phys. Soc. Jpn* **83** 063704

[28] Fujioka M, Nagao M, Denholme S J, Tanaka M, Takeya H, Yamaguchi T and Takano Y 2014 High-Tc Phase of PrO0.5F0.5BiS2 single crystal induced by uniaxial pressure *Appl. Phys. Lett.* **105** 052601

[29] Fujioka M, Tanaka M, Denholme S J, Yamaki T, Takeya H, Yamaguchi T and Takano Y 2014 Pressure-induced phase transition for single-crystalline LaO0.5F0.5BiSe2 *EPL* **108** 47007

[30] Hisashi K, Tomita Y, Tou H, Izawa H, Mizuguchi Y, Miura O, Demura S, Deguchi K and Takano Y 2012 Pressure Study of BiS2 -Based Superconductors Bi4O4S3 and La(O,F)BiS2 *J. Phys. Soc. Jpn* **81** 103702

[31] Chu C W, Hor P H, Meng R L, Gao L and Huang Z J 1987 superconductivity at 52.5 K in Lanthanum-Barium-Copper-Oxide System *Science (80-. ).* **235** 567–9

[32] Gao L, Xue Y Y, Chen F, Xiong Q, Meng R L, Ramirez D, Chu C W, Eggert J H and Mao H K 1994 Superconductivity up to 164 K in HgBa2Cam-1Cum02m+2+x (m=1, 2, and 3) under quasihydrostatic pressures *Phys Rev B* **50** 4260–3

[33] Jha R, Kishan H and Awana V P S 2014 Effect of hydrostatic pressures on the superconductivity of new BiS2 based REO0.5F0.5BiS2 (RE=La, Pr and Nd) superconductors *J. Phys. Chem. Solids* **84** 17–23

[34] Wolowiec C T, White B D, Jeon I, Yazici D, Huang K and Maple M B 2013 Enhancement of superconductivity near the pressure-induced semiconductor-metal transition in *J. physics. Condens. matter* **25** 422201

[35] Honda F, Kaji S, Minamitake I, Ohashi M, Oomi G, Eto T and Kagayama T 2002 High-pressure apparatus for the measurement of thermal and transport properties at multi-extreme conditions *J. Phys.: Condens. matter* **14** 11501–5

[36] Fang Y, Yazici D, White B D and Maple M B 2015 Enhancement of superconductivity in La1−xSmxO0.5F0.5BiS2 *Phys. Rev. B* **91** 1–6

[37] Kajitani J, Omachi A, Hiroi T, Miura O and Mizuguchi Y 2014 Chemical pressure effect on Tc in BiS2 - based Ce1-xNdxO0.5F0.5BiS2 *Physica C* **504** 33–5

[38] Pallechi I, Lamura G, Putti M, Kajitani J, Y. Mizuguchi, Miura O, Demura S, Deguchi K and Takano Y 2014 Effect of high pressure annealing on the normal state transport of LaO0.5F0.5BiS2 *Phys Rev B* **89** 214513

[39] Jha R, Tiwari B and Awana V P S 2015 Appearance of bulk superconductivity under hydrostatic pressure in( RE = Ce , Nd , Pr , and Sm ) compounds *J. Appl. Phys.* **117** 013901

[40] Morice C, Artacho E, Dutton S E, Molnar D, Kim H and Saxena S S 2013 Effects of stoichiometric doping and defect layers in superconducting Bi-O-S compounds *J. Phys. Condens. Matter* **27** 135501

[41] Morice C, Artacho E, Dutton S E, Molnar D, Kim H and Saxena S S Electronic and magnetic properties of superconducting Ln O1−xFxBiS2 (Ln = La, Ce, Pr and Nd) from first principles *arXiv:1312.2615v1*

[42] Shao J, Liu Z, Yao X, Pi L, Tan S, Zhang C and Zhang Y 2014 Bulk superconductivity in single *Phys. status solidi - Rapid Res. Lett.* **8** 845–8

[43] Benayad N, Djermouni M and Zaoui A 2014 Computational Condensed Matter Electronic structure of new superconductor La0.5Th0.5OBiS2 : DFT study *Comput. Condens. matter* **1** 19–25

**Figure captions**

**Figure.1. (color online)** Crystal Structure of layered LnOBiS$_2$ materials

**Figure.2. (color online)** The dc magnetization recorded at 10 Oe as a function of temperature at different applied pressures for La$_{1-x}$Sm$_x$O$_{0.5}$F$_{0.5}$BiS$_2$ (a) $x = 0.2$ and (b) $x = 0.8$ in the low temperature range. Inset of the both figures show the change in $T_c$ as a function of pressure in the magnetization measurements.

**Figure.3**. **(color online)** Temperature dependence of the resistivity of La$_{1-x}$Sm$_x$O$_{0.5}$F$_{0.5}$BiS$_2$ at various pressures (a and b) $x = 0.2$ and (c and d) $x = 0.8$ and variation of magnetic and resistive $T_c$ with pressure (e and f). Insets of (a and c) shows the superconducting transition for $x = 0.2$ and 0.8 at ambient pressure.

**Figure.4. (color online)** log($\rho$) vs 1/T plot at various pressures for La$_{1-x}$Sm$_x$O$_{0.5}$F$_{0.5}$BiS$_2$ (a) $x = 0.2$ and (b) $x = 0.8$. The vertical and horizontal solid lines are the linear fitting to the activation energy equation at high and low temperature respectively.

**Figure.5. (color online)** Pressure dependences of $\rho$ (at 11 K, just above $T_c$), RRR and calculated energy gaps ($\Delta_1$ and $\Delta_2$) for La$_{1-x}$Sm$_x$O$_{0.5}$F$_{0.5}$BiS$_2$ (a and b) $x = 0.2$ and (c and d) $x = 0.8$.

**Figures**

**Figure.1.**

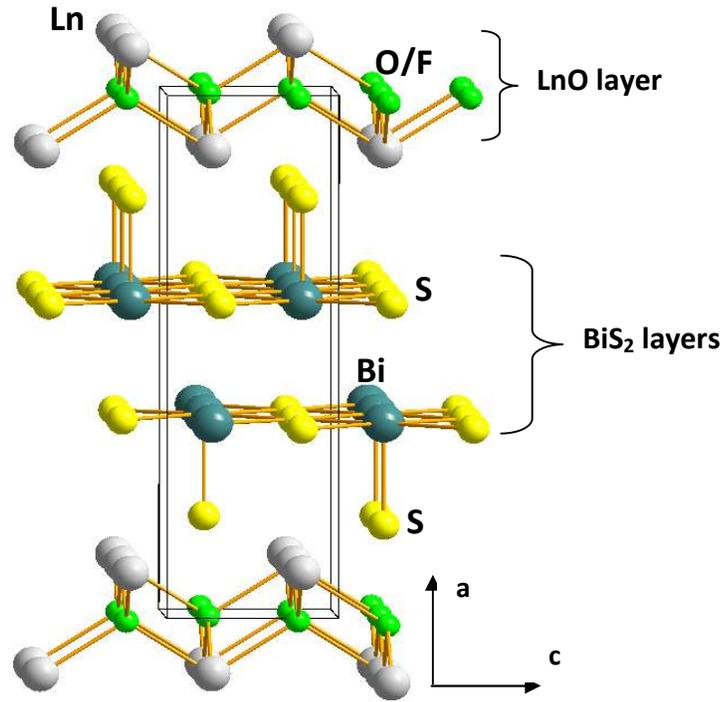

**Figure.2.**

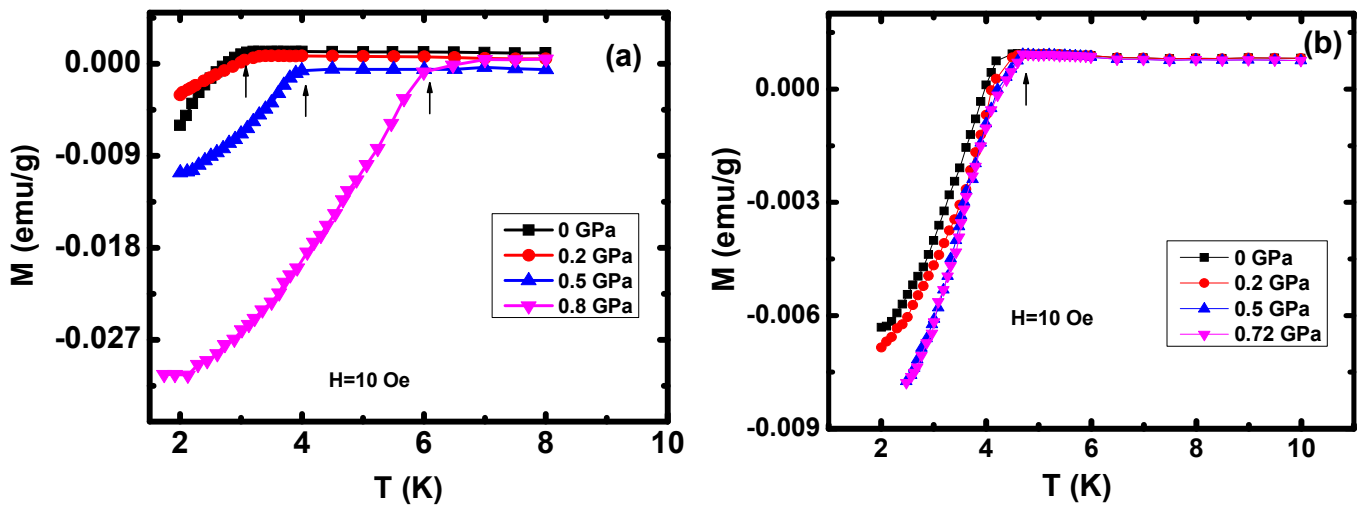

**Figures**

**Figure.1.**

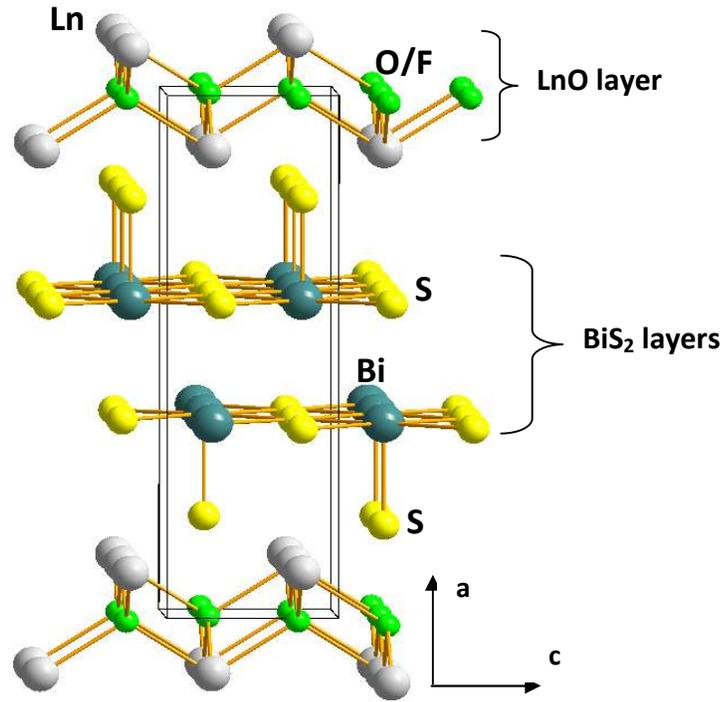

**Figure.2.**

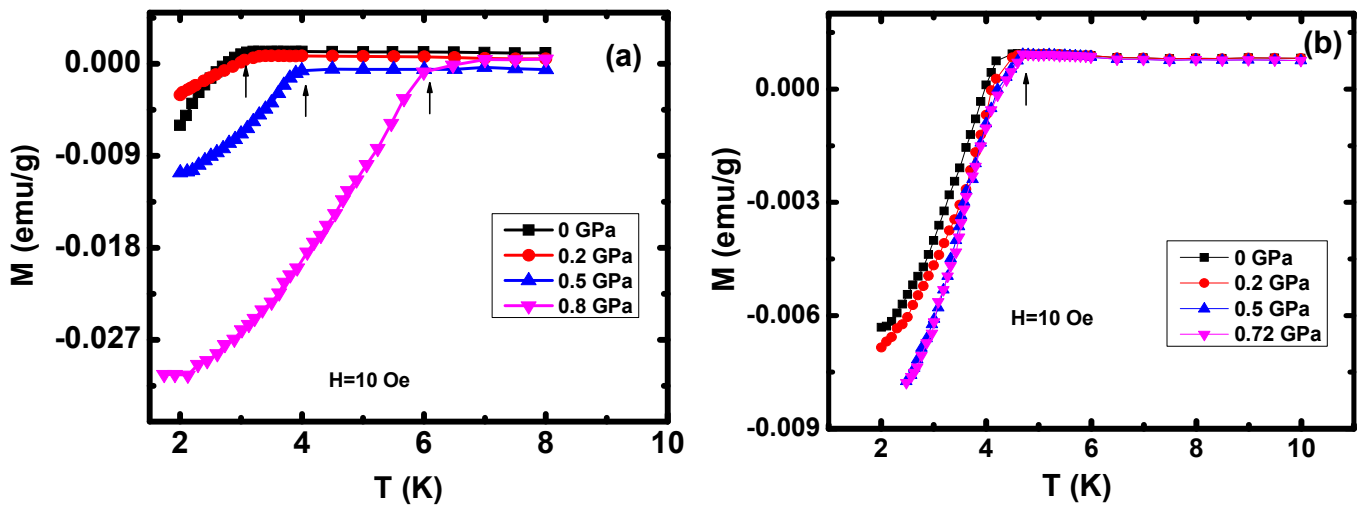

**Figure. 3.**

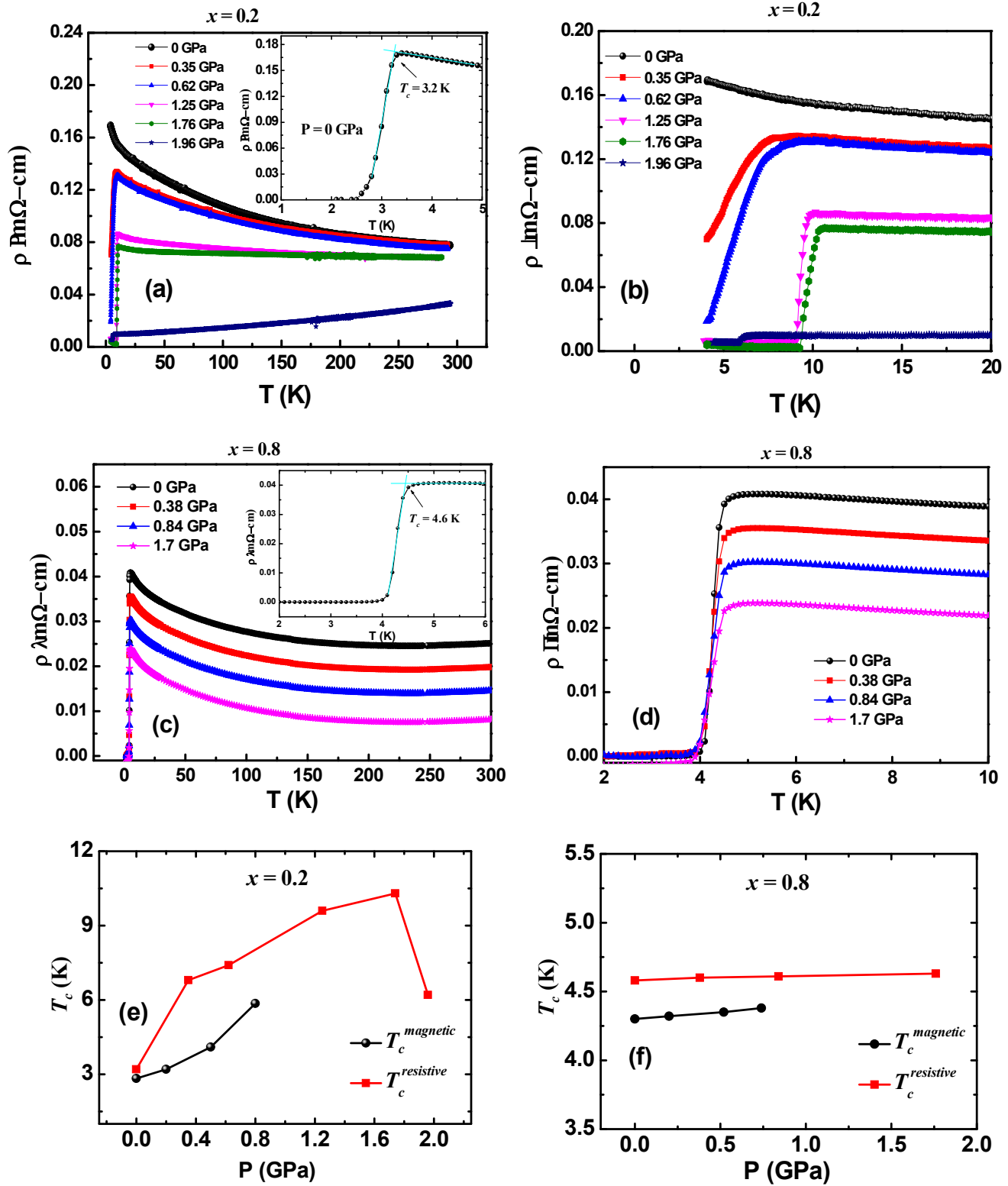

**Figure. 4.**

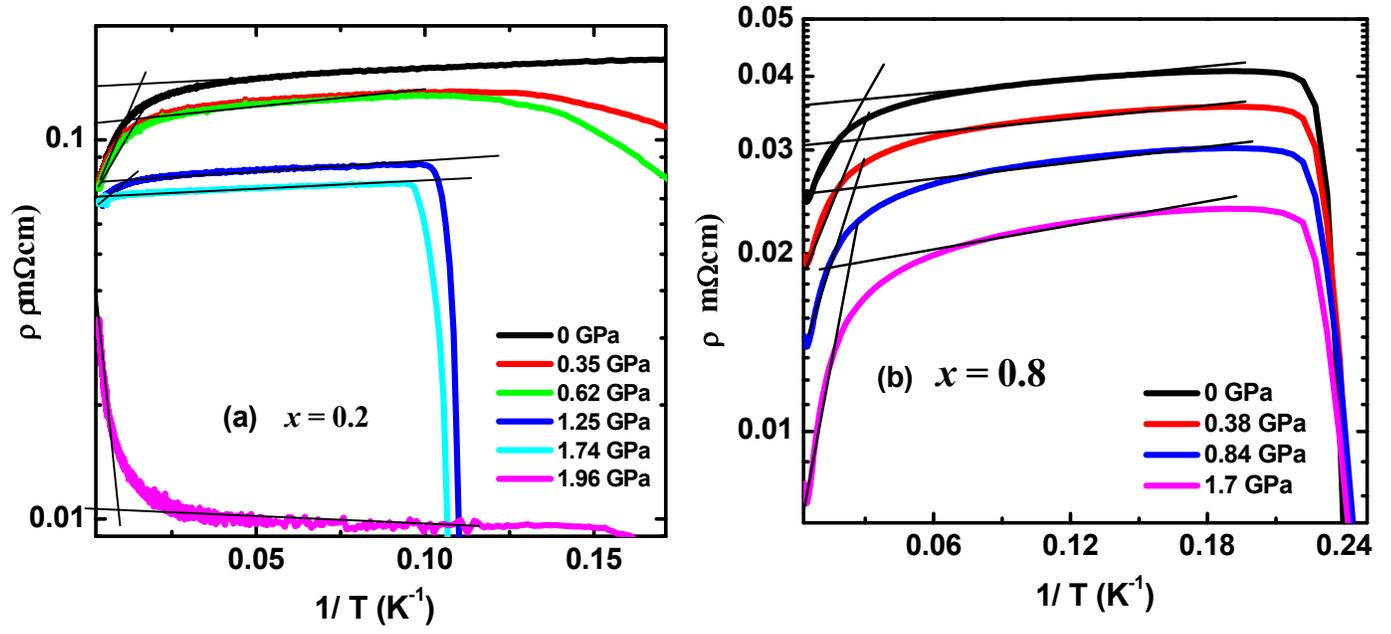

**Figure. 5.**

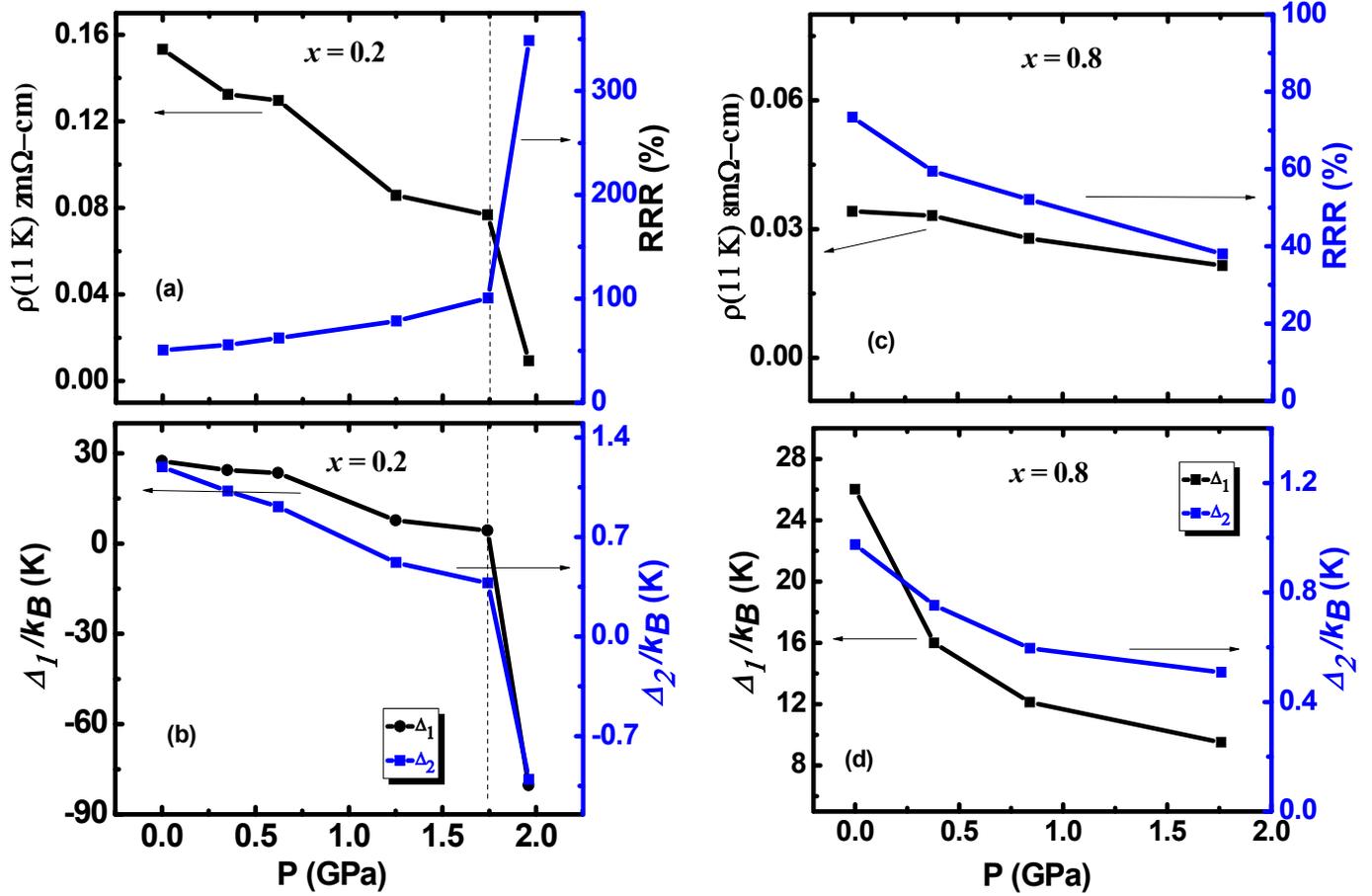